\documentclass[onecolumn]{article}
\usepackage[english]{babel}
\usepackage{amssymb,amsfonts,mathrsfs}
\usepackage[dvips]{graphicx}

\begin{document}

\newcommand{\vo}{\vec{v}_{(1)}}
\newcommand{\vj}{\vec{v}_{(j)}}
\newcommand{\Hilb}{\mathscr{H}}
\newcommand{\I}{\mathbb{I}}
\newcommand{\n}{\hat{n}}
\newcommand{\ab}[1]{a^{\phantom{\dag}}_{#1}}
\newcommand{\ad}[1]{a^{\dag}_{#1}}
\newcommand{\LOP}[1]{\underline{#1}}
\newcommand{\LOPF}[1]{\overline{#1}}

\title{ \textbf{ Exploring a New Post-Selection Condition for Efficient Linear Optics Quantum 	
Computation}}
\date{}

\maketitle

\begin{center}
	\author{R. Coen Cagli\footnote{email: $\mathtt{ruben.coencagli@na.infn.it}$}$^{,a}$, P. Aniello$^{a,b}$, N.Cesario$^{c}$, F. Foncellino$^{c}$\\
		\small \textit{a. Dipartimento di Fisica dell'Universit\`{a} di Napoli Federico II,\\
		Complesso Univ. M.S. Angelo via Cintia, Napoli, 80126, Italy.}\\
		\textit{b. Istituto Nazionale di Fisica Nucleare (INFN), Sez. di Napoli.}\\
		\textit{c. SST Corporate R\&D STMicroelectronics,\\ via Remo De Feo,1, Arzano(NA), 80022, 			Italy.}
	}
\end{center}


\begin{abstract}
Recently, it was shown that fundamental gates for theoretically
efficient quantum information processing can be realized by using
single photon sources, linear optics and quantum counters. One of
these fundamental gates is the NS-gate, that is, the one-mode
non-linear sign shift. In this work, firstly, we prove by a new
rigorous proof that the upper bound of success probability of
NS-gates with only one helper photon and an undefined number of
ancillary modes is bounded by $0.25$. Secondly, we explore the
upper bound of success probability of NS-gate with a new
post-selection measurement. The idea behind this new
post-selection measurement is to condition the success of NS-gate
transformation to the observation of only one helper photon in
whichever of the output modes.
\end{abstract}

\section{Introduction to Conditional Operations}

Linear optical \emph{passive} (LOP) transformations are defined as
the class of linear optical transformations that act on the system
of $N$ optical modes leaving unchanged the total number of photons
in the process. With every LOP one can associate three
mathematical objects: the $N \times N$ unitary matrix
$\LOP{U}$ describing the transformation on the field
operators of the $N$ optical modes, the unitary operator $U$
acting by similarity on the field operators, and the unitary
infinite matrix $\LOPF{U}$ representing $U$ on the Fock space
of the optical modes.

In the context of quantum computing and quantum information
processing, several conditional schemes have been
proposed~\cite{klm1,pjf1,ral1,ral2,ral3,ral4,giorgi} to perform a wider
class of transformations. It is still an open problem the complete
classification of this wider class~\cite{scheel,lapaire,clausen}.
However, the general scheme has a conceptually simple two step
structure. At first, one couples the $N$ mode system with $k$
\emph{ancillary} modes and let the two transform  under a global
LOP, corresponding to a unitary evolution of the bipartite system.
Then one performs a measurement on the ancillae and selects the
output state of the system only when a  predefined result is
obtained: thisi is called \emph{post selection}. This procedure in
general will transform the state of the system as a
\emph{completely positive} map (CP).

These schemes are referred to as \emph{conditional} because the
implemented transformations are conditioned by a predefined
measurement outcome. Moreover, these schemes are referred to as
\emph{non-deterministic} because there is some probability that a
different outcome is obtained (i.e. the schemes implement a
transformation that is not the desired one).

We start introducing some notation: let us denote with $\Hilb_{S}$
the Hilbert space on which the logical system is encoded. Indeed,
logical states can be encoded in a subspace of the $N$ mode Fock
space or more in general in the direct sum of several such
subspaces. So we can write: $\Hilb_{S} =
\bigoplus_{n}\Hilb_{n_{S}}^{(N)}$, where we denote by
$\Hilb_{n_{S}}^{(N)}$ the space spanned by the states of $n_{S}$
photons distributed on the $N$ modes.

Let $\rho$ be the input state of the system, i.e. a density matrix
on $\Hilb_{S}$, and $\sigma$ the ancilla input state, and we
assume that $\sigma$ is a pure Fock state with exactly $n_{A}$
photons, i.e. a rank $1$ projector on the Hilbert space:
$\Hilb_{A} = \Hilb_{n_{A}}^{(k)}$. If the ancilla and the system
do not interact during their preparation, the global input state
is not an entangled one, so that we can write it as:
$\rho\otimes\sigma$.

The effect on the system of the global LOP can be described by the
expression $U(\rho\otimes\sigma) U^{\dag}$, which corresponds a
mixed output with respect to the system S:
$\rho^{'}=\textrm{tr}_{A}(U(\rho\otimes\sigma) U^{\dag})$. It is
well known that this transformation can be described by a trace
preserving CP map $\tau$ for which one can always find a
\emph{operator sum representation}:
\begin{eqnarray}\label{10}
\rho^{'} = \tau(\rho) = \sum_{\mu}M_{\mu}\rho M_{\mu}^{\dag} \nonumber \\
 \sum_{\mu}M_{\mu}^{\dag}M_{\mu} = \I
\end{eqnarray}
\noindent After a suitable relabeling of $\Hilb_{S}$ and
$\Hilb_{A}$, denoting with $\{|\alpha\rangle_{S}\}$ and
$\{|\nu\rangle_{A}\}$ their respective basis, $\rho$ and $\sigma$
can be decomposed as:
\begin{equation}\label{11}
\left\{\begin{array}{ccc}
\rho & = & \sum_{\alpha,\beta}\rho_{\alpha\beta}|\alpha\rangle_{S}\langle \beta| \\
\sigma & = & |\nu\rangle_{A}\langle\nu|
\end{array}\right.
\end{equation}
\noindent Using these expressions we can explicitly evaluate the
partial trace obtaining the output density matrix of the system S,
\begin{eqnarray}\label{12}
\rho^{'}_{\gamma\delta}  & = &
\sum_{\mu}\sum_{\alpha,\beta}(\langle\mu^{'}|\otimes\langle
\gamma^{'}|U|\alpha\rangle\otimes|\nu\rangle)
\nonumber\\
 {} & {} & \rho_{\alpha\beta}(\langle\nu|\otimes\langle \beta|U^{\dag}|\delta^{'}\rangle\otimes|\mu^{'}\rangle)
\end{eqnarray}
\noindent and so the matrix elements of $M_{\mu}$ are given by:
\begin{equation}\label{13}
(M_{\mu}^{(\nu)})_{\gamma}^{\alpha}  =
\langle\mu^{'}|\otimes\langle
\gamma^{'}|U|\alpha\rangle\otimes|\nu\rangle
\end{equation}
\noindent where index $\nu$ is fixed by the ancilla input, index
$\mu$ is related to the post-selection condition, and
$\gamma,\alpha$ run through matrix elements. It is easy to verify
that unitarity of $U$ guarantees condition \ref{10}.

We consider the case in which this is described by a
\emph{Projective Valued Measure} (PVM) associated to the basis $\{
|\mu^{'}\rangle\} $, namely by rank $r$ projectors:
\begin{equation}\label{14}
  P_{\mu}  = \I_{S}\otimes\sum_{\mu}s_{\mu}|\mu^{'}\rangle\langle\mu^{'}|
\end{equation}
\noindent with exactly $r$ terms in which $s_{\mu}=1$, and all the
others with $s_{\mu}=0$.

Now, the conditional (unnormalized) output state is:
\begin{equation}\label{15}
\bar{\rho}  = \textrm{tr}_{A}(P_{\mu}U(\rho\otimes\sigma)
U^{\dag}P_{\mu}) = M_{\mu}\rho M_{\mu}^{\dag}
\end{equation}
\noindent with probability
\begin{equation}\label{16}
p_{\mu}  = \textrm{tr}_{S}\textrm{tr}_{A}(U(\rho\otimes\sigma)
U^{\dag} P_{\mu}) =  \textrm{tr}_{\bar{S}}(M_{\mu}\rho
M_{\mu}^{\dag})
\end{equation}
\noindent Of course the normalized output state is given by
$\bar{\rho}_{\textrm{\footnotesize{norm}}} =
\frac{\bar{\rho}}{p_{\mu}}$.

\section{Non-Linear Sign-Shift Gate}

\subsection{Gate Operation}

Now, we are interested in analyzing the implementation of the
one-mode non-linear sign-shift (NS) on the 2-photons Fock state:
\begin{equation}\label{18}
|\psi\rangle = \alpha|0\rangle + \beta|1\rangle + \gamma|2\rangle
\longrightarrow |\psi^{'}\rangle = \alpha|0\rangle +
\beta|1\rangle - \gamma|2\rangle
\end{equation}
\noindent as a conditional operation. This transformation is not
realizable as a one-mode LOP (hence the name non-linear). So we
consider the conditional scheme proposed by KLM uses two ancillary
modes prepared in the state, $\sigma = |10\rangle_{A}\langle 10|$,
and the post-selection condition described by the rank-1
projector, $P_{10}  = \I_{S}\otimes |10\rangle_{A}\langle 10|$.
The corresponding non-unitary operator is represented by the
matrix:
\begin{equation}\label{22}
(M_{10}^{(10)})_{n^{'}}^{n}  =  {}_{A}\langle 10|\otimes
{}_{S}\langle n^{'}|U|n\rangle_{S}\otimes |10 \rangle_{A}
\end{equation}
\noindent where $ |n\rangle_{S}$ is the $n$-photon Fock state. The
conservation of the total number of photons implies that
$M_{10}^{(10)}$ should be a diagonal matrix, since non-vanishing
terms are those with $n=n^{'}$, and by a straightforward
calculation one finds:
\begin{eqnarray}\label{23}
(M_{10}^{(10)})_{0}^{0}  =  & \LOPF{U}_{010}^{010} & = \LOP{U}_{22} \nonumber \\
(M_{10}^{(10)})_{1}^{1}  =  & \LOPF{U}_{110}^{110}  & = \LOP{U}_{11}\LOP{U}_{22} + \LOP{U}_{12}\LOP{U}_{21} \nonumber \\
(M_{10}^{(10)})_{2}^{2}  =  & \LOPF{U}_{210}^{210}  & =
\LOP{U}_{11}(\LOP{U}_{11}\LOP{U}_{22} + 2\LOP{U}_{12}\LOP{U}_{21})
\end{eqnarray}
\noindent For the operation to implement the desired sign-shift,
one must ask that $\LOP{U}$ is such that:
\begin{equation}\label{24}
(M_{10}^{(10)})_{0}^{0} = (M_{10}^{(10)})_{1}^{1} =
-(M_{10}^{(10)})_{2}^{2}
\end{equation}
\noindent which means that,
\begin{eqnarray}\label{26}
\LOP{U}_{11} & = & 1-\sqrt{2} \nonumber \\
\LOP{U}_{22} & = & \frac{ \LOP{U}_{12} \LOP{U}_{21}}{1-
\LOP{U}_{11}}
\end{eqnarray}
\noindent The success  probability is obtained simply applying
(\ref{16}):
\begin{equation}\label{25}
p_{10} = |(M_{10}^{(10)})_{0}^{0}|^{2}   = |\LOP{U}_{22}|^{2}
\end{equation}
\noindent and it is maximum when
$\LOP{U}_{12}=\LOP{U}_{21}=2^{-\frac{1}{4}}$, which gives $p_{10}
= \frac{1}{4}$.

Two possible circuits implementing this operation are shown
in~\cite{klm1,ral1}.

\subsection{General Bound with One Ancillary Photon}

In this section we show that the maximum value for $p_{\mu}$ when
only one ancillary photon is present at the input cannot be
increased by adding any arbitrary number of ancillary modes
prepared in the vacuum state.
\footnote{ Different approaches to  the
exploration of the upper bound of success probability of NS-gate
can be found in recent works ~\cite{knipro,knill03,scheelpro}.}

Let us suppose that we have a $k$-modes initial ancillary state,
$\sigma_{i} = |\mathbf{i}\rangle_{A}\langle \mathbf{i}|$, where
$|\mathbf{i}\rangle_{A}$ is the state with one photon injected in
the $i$-th mode, and a post-selection condition, $P_{j}  =
\I_{S}\otimes |\mathbf{j}\rangle_{A}\langle \mathbf{j}|$.

\noindent It should be clear that whichever input state
$\sigma_{l}$ is related to $\sigma_{i}$ by a simple permutation,
namely a swapping of two modes that can be done deterministically,
and the same holds for any post-selection condition $P_{l}$. From
a mathematical viewpoint, it simply consists in the exchange of
two rows, or two columns of $\LOP{U}$.

We first impose the functioning conditions, and then study the
probability under the request that the matrix $\LOP{U}$ should be
unitary in order to be implementable by a LOP circuit. It is
simple to show that the $(k+1)\times(k+1)$ matrix $\LOP{U}$ must
satisfy the same conditions as in (\ref{26}), but with column
index $2$ replaced by $i$ and row index $2$ replaced by $j$. Thus
one can write:
\begin{equation}\label{30}
\left( \begin{array}{cccc}  1- \sqrt{2} & \ldots & \LOP{U}_{1,i} & \ldots\\
                            \vdots & {} & \vdots & {}\\
             \LOP{U}_{j,1} & \ldots & \frac{\LOP{U}_{1,i}\LOP{U}_{j,1}}{\sqrt{2}} & \ldots \\
             \vdots & {} & \vdots & \ddots
\end{array} \right)
\end{equation}
\noindent Now, the probability is equal to
$\frac{\LOP{U}_{1,i}\LOP{U}_{j,1}}{\sqrt{2}}$, and it has to be
maximized under the condition that $\LOP{U}$ be unitary. We note
that it suffices to impose that the first and the $j$-th row are
mutually orthogonal, and that they are normalizable together with
the first and the $i$-th column, namely:
\begin{equation}\label{31}
\sum_{l}\LOP{U}_{1,l} = 1 = \sum_{l}\LOP{U}_{j,l}  \quad
\sum_{l}\LOP{U}_{l,1} = 1 = \sum_{l}\LOP{U}_{l,i} \quad
\end{equation}
\noindent When one has two orthonormal rows, the whole matrix can
be constructed simply by completing the set of orthonormal vectors
arbitrarily, since all other matrix elements do not enter into the
functioning conditions. Normalizability is expressed by:
\begin{eqnarray}\label{32a}
\left\{ \begin{array}{ccc} \textrm{row} \quad 1  & \rightarrow &
|U_{1,i}|^{2}\leq 2(\sqrt{2}-1) \nonumber \\
\textrm{row} \quad j & \rightarrow & |U_{j,1}|^{2}\leq
\left(1+\frac{|U_{1,i}|^{2}}{2}\right)^{-1} \nonumber \\
\textrm{column} \quad 1 & \rightarrow & |U_{j,1}|^{2}\leq
2(\sqrt{2}-1) \nonumber \\
\textrm{column} \quad i & \rightarrow & |U_{1,i}|^{2}\leq
\left(1+\frac{|U_{j,1}|^{2}}{2}\right)^{-1}
\end{array}\right.
\end{eqnarray}

\noindent and this furnishes a limitation for the region in which
$|U_{1,i}|$ and $|U_{j,1}|$ can take values (see
fig.1).

\noindent To impose orthogonality of two rows of arbitrary length,
we define the two $(k-1)$-components complex vectors $\vo,\vj$:
\begin{eqnarray}\label{32}
\vo^{\top} & = & (U_{1,2},\ldots,U_{1,i-1},U_{1,i+1},\ldots,U_{1,k+1})  \\
\vj^{\top} & = &
(U_{j,2},\ldots,U_{j,i-1},U_{j,i+1},\ldots,U_{j,k+1})
\end{eqnarray}
\noindent such that normalization of the rows is completed
\begin{equation}\label{33}
\left\{ \begin{array}{c}
|\vo|^{2}= 2(\sqrt{2}-1)-|U_{1,i}|^{2} \\
|\vj|^{2} = 1-|U_{j,1}|^{2}\left(1+\frac{|U_{1,i}|^{2}}{2}\right)
\end{array}\right.
\end{equation}
\noindent and orthogonality is satisfied
\begin{equation}\label{34}
\vo^{\dag}\cdot\vj =
-U_{j,1}\left(1-\sqrt{2}+\frac{|U_{1,i}|^{2}}{\sqrt{2}}\right)
\end{equation}
\noindent Here  $\vo^{\dag}$ denotes the  hermitian conjugate of
$\vo$, namely the row whose elements are the complex conjugate of
those of $\vo$. Now, we notice that in a complex vector space,
making use of the Schwartz inequality, the scalar product can be
written as:
\begin{equation}\label{35}
\vo^{\dag}\cdot\vj = e^{i\phi}|\vo||\vj|\cos\alpha_{1j}
\end{equation}
\noindent and upon substitution $U_{j,1}=e^{i\phi_{j1}}|U_{j,1}|$,
we can separate eq. (\ref{34}) in phase and modulus:
\begin{equation}\label{36}
\left\{ \begin{array}{c}
\phi = \phi_{j1} \pm \pi\\
|\vo^{\dag}\cdot\vj| =
|U_{j,1}|\left|\left(1-\sqrt{2}+\frac{|U_{1,i}|^{2}}{\sqrt{2}}\right)\right|
\end{array} \right.
\end{equation}
\noindent After some substitutions, one gets:
\begin{equation}\label{37}
|\cos\alpha_{1j}| =
\frac{|U_{j,1}||(1-\sqrt{2})+\frac{|U_{1,i}|^{2}}{\sqrt{2}}|}
{\sqrt{(2(\sqrt{2}-1)-|U_{1,i}|^{2})(1-|U_{j,1}|^{2}(1+\frac{|U_{1,i}|^{2}}{2}))}}
\end{equation}
\noindent this means that the required vectors $\vo,\vj$ exist
only within the region where the r.h.s. of (\ref{37}) is bounded
by $1$. To simplify the notation, we make the following
substitutions:
\begin{eqnarray}\label{38}
x \doteq |U_{1,i}| \quad y \doteq |U_{j,1}| \\
\left\{ \begin{array}{c}
A = |(1-\sqrt{2})+\frac{x^{2}}{\sqrt{2}}|\\
B = 2(\sqrt{2}-1)-x^{2}\\
C = 1+\frac{x^{2}}{2} \end{array} \right.
\end{eqnarray}
\noindent We find that the l.h.s. in (\ref{37}) takes its maximum
acceptable value, namely $1$, on the boundary of the region
depicted in fig. 1,
\begin{figure}[!h]
\begin{center}\label{figura1}
\includegraphics[width=0.55\columnwidth]{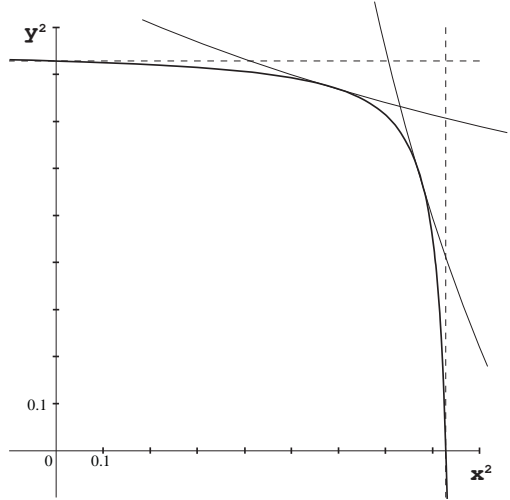}
\caption{Domain limitations for $x^{2}$ and $y^{2}$.}
\end{center}
\end{figure}
\noindent which is described by the following equation,
\begin{equation}\label{39}
y^{2} =  \frac{B}{A^{2}+BC} =  \frac{2(\sqrt{2}-1)-x^{2}}
{|(1-\sqrt{2})+\frac{x^{2}}{\sqrt{2}}|^{2}+(2(\sqrt{2}-1)-x^{2})(1+\frac{x^{2}}{2})}
\end{equation}
\noindent Now the problem of maximizing the probability $p_{j}$
becomes trivial, because  $p_{j} = \frac{x^{2}y^{2}}{2}$ and one
can simply substitute $y^{2}$ with the value it takes along the
curve (\ref{39}), and maximize $p_{j}$ as a function of only one
variable $x^{2}$,
\begin{equation}\label{41}
p_{j} = \frac{x^{2}}{2}\cdot\frac{2(\sqrt{2}-1)-x^{2}}
{|(1-\sqrt{2})+\frac{x^{2}}{\sqrt{2}}|^{2}+(2(\sqrt{2}-1)-x^{2})(1+\frac{x^{2}}{2})}
\end{equation}
\noindent This takes its maximum value in
$x^{2}=y^{2}=\frac{1}{\sqrt{2}}$, which is (see fig. 2), $p_{j} = 0.25$.
\begin{figure}[!h]
\begin{center}\label{figure2}
\includegraphics[width=0.55\columnwidth]{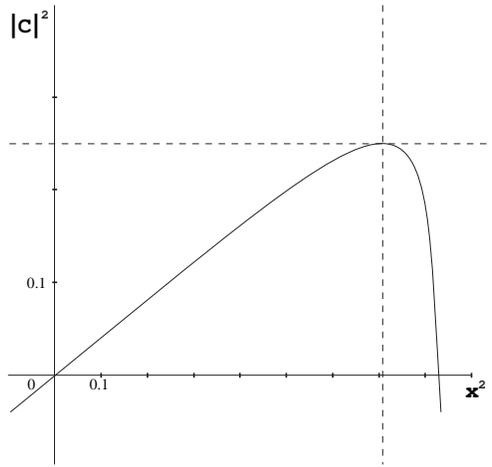}
\caption{Limitations on success probability of NS-gate: the
maximum value is $0.25$.}
\end{center}
\end{figure}
%

\subsection{Generalizing the Input State}

When adding $k$ ancillary modes to the system, we are considering
a much more general situation than that described by input
$\sigma_{i}$: actually, we are considering the simplest case in
which the ancillary Hilbert space is enlarged to an arbitrary
dimensionality $k$. This is because the state
$|\mathbf{i}\rangle_{A}$ can be transformed in reversible
deterministic way to any normalized one-photon state,
$|\chi\rangle_{A} = \sum_{i=2}^{k+1} \gamma_{i}\ad{i}|0\ldots
0\rangle_{A}$, by a LOP acting only on the $k$ ancillary modes,
that we denote by $V=\I_{S}\otimes V_{A}$. Therefore, any circuit
acting on $|\psi\rangle_{S}\otimes |\chi\rangle_{A}$ as $U$  can
be reduced to one acting on $|\psi\rangle_{S}\otimes
|\mathbf{i}\rangle_{A}$ denoted by $U^{'}=UV$,
\begin{equation}\label{44}
U(|\psi\rangle_{S}\otimes |\chi\rangle_{A}) =
U(V(|\psi\rangle_{S}\otimes |\mathbf{i}))=
U^{'}(|\psi\rangle_{S}\otimes |\mathbf{i}\rangle_{A})
\end{equation}
\noindent and the result of the previous section is still valid
with general one-photon input state.

\subsection{Generalizing Post-Selection Condition}

Further generalization is obtained considering the possibility of
implementing the following transformation,
\begin{equation}\label{45}
 |\psi\rangle_{S}\otimes |\mathbf{i}\rangle_{A} \longrightarrow
|\psi^{'}\rangle_{S}\otimes \sum_{j=2}^{s+1}
\gamma_{j}\ad{j}|0\ldots 0\rangle_{A}
\end{equation}
\noindent Notice that here the ancillary state is not normalized,
due to non-unit probability of success. Even if in this case
amplitudes $\gamma_{j}$ cannot be summed, every time the photon is
observed in the $j$-th output mode the desired NS-gate transform
is obtained, and this happens with probability
$p_{j}=|\gamma_{j}|^{2}$.

\noindent We are in the  situation of a rank-$s$ post-selection
condition,
\begin{equation}\label{46}
P_{s}  = \I_{S}\otimes
\sum_{j=2}^{s+1}|\mathbf{j}\rangle_{A}\langle \mathbf{j}|
\end{equation}
\noindent corresponding to the observation of only one photon in
\emph{whichever} of the output modes. Then the total probability
of success is, $p_{\textrm{\small{tot}}} = \sum_{j}p_{j}$.

\noindent As the simplest example, one could consider the case
where $\sigma=|10\rangle_{A}\langle 10|$ and $P=\I_{S}\otimes
(|10\rangle_{A}\langle 10|+|01\rangle_{A}\langle 01|)$, and find
the following equations for the gate functioning,
\begin{eqnarray}\label{48}
\LOP{U}_{11} & = & 1-\sqrt{2} \nonumber \\
\LOP{U}_{22} & = & \frac{ \LOP{U}_{12} \LOP{U}_{21}}{1- \LOP{U}_{11}}\nonumber \\
\LOP{U}_{32} & = & \frac{ \LOP{U}_{12} \LOP{U}_{31}}{1-
\LOP{U}_{11}}
\end{eqnarray}
\noindent so that
\begin{equation}\label{49}
p_{\textrm{\small{tot}}}  =
\frac{|\LOP{U}_{12}|^{2}}{2}(|\LOP{U}_{21}|^{2}|\LOP{U}_{31}|^{2})
\end{equation}
\noindent In general, one has $k=s+m$ ancillary modes, being the
photon injected in mode $i\leq s$. The additional $m$ modes are
needed to guarantee that the circuit can be implemented as a LOP,
namely to make the $\LOP{U}$ matrix indeed unitary.

Functioning conditions constrain $\LOP{U}$ to be in the form:
\begin{equation}\label{50}
\left(\begin{array}{cccc}  1- \sqrt{2} & \ldots & \LOP{U}_{1,i} & \ldots\\
\LOP{U}_{2,1} & \ldots & \frac{\LOP{U}_{2,1}\LOP{U}_{1,i}}{\sqrt{2}} & \ldots\\
\vdots & {} & \vdots & {} \\
\LOP{U}_{s+1,1} & \ldots & \frac{\LOP{U}_{s+1,1}U_{1,i}}{\sqrt{2}} & \ldots \\
\LOP{U}_{s+2,1} & \ldots & \LOP{U}_{s+2,i} & \ldots \\
\vdots & {} & \vdots & \ddots
\end{array} \right)
\end{equation}
\noindent In this case, the success probability of NS-gate has the
following form:
\begin{equation}\label{51}
p_{\textrm{\small{tot}}}  =
\frac{|\LOP{U}_{1i}|^{2}}{2}\sum_{j=2}^{s+1}|\LOP{U}_{j1}|^{2}
\end{equation}
\noindent Following the lines of the subsection 2.2 one finds that
all the calculations are still valid in this case, and
$p_{\textrm{\small{tot}}}$ has to be maximized along the curve in
fig. 2 with the replacement
$x^{2}=\sum_{j=2}^{s+1}|\LOP{U}_{j1}|^{2}$. The result is that the
same upper bound holds, namely $p_{\textrm{\small{tot}}}\leq
0.25$.

\noindent One could have also argued this result, by observing
that the global LOP transformation in this case gives:
\begin{eqnarray}\label{52}
U(|\psi\rangle_{S}\otimes |\mathbf{i}\rangle_{A}) & \!\!\!\! = \!\!\!\!&
|\psi_{+}\rangle_{S}\otimes
 \gamma_{0}|0\ldots 0\rangle_{A} + |\psi^{'}\rangle_{S}\otimes
 \sum_{l=1}^{s+1}\gamma_{l}\ad{l}|0\ldots 0\rangle_{A} +  \nonumber \\
{} & \!\!\!\!\!\!\!\!\!\!\!\! \!\!\!\!\!\!\!\!\!\!\!\!\!\!\!\!\!\!\!\! \!\!\!\!\!\!\!\!\!\!\!\!\!\!\!\!\!\!\!\! 
\!\!\!\!\!\!\!\!\!\!\!\!\!\!\!\!\!\!\!\! \!\!\!\!\!\!\!\!\!\!\!\!\!\!\!\!\!\!\!\! \!\!\!\!\!\!\!\!
 + & \!\!\!\!\!\!\!\!\!\!\!\! \!\!\!\!\!\!\!\!\!\!\!\! \!\!\!\!\!\!\!\!\!\!\!\!\!\!\!\!\!\!\!\!\!\!\!\! \!\! |\psi_{-}\rangle_{S}\otimes
 \sum_{l,m=1}^{s+1}\gamma_{lm}\ad{l}\ad{m}|0\ldots 0\rangle_{A} +  |\psi_{--}\rangle_{S}\otimes
 \sum_{l,m,n=1}^{s+1}\gamma_{lmn}\ad{l}\ad{m}\ad{n}|0\ldots
 0\rangle_{A}
\end{eqnarray}
\noindent where indices $+,-$ denote states of the system S with
one photon added or subtracted, and
$|\psi_{--}\rangle_{S}=|0\rangle_{S}$ necessarily.

The point here is that a subsequent LOP $V$ acting only on modes
from $2$ to $s$ before post-selection measurement would leave
invariant the subspaces of the ancillary Fock space with any fixed
number of photons. Thus, one can always find a suitable $V$ that
maps reversibly and deterministically the state (\ref{52})  to
another one of the form:
\begin{eqnarray}\label{53}
U(|\psi\rangle_{S}\otimes |\mathbf{i}\rangle_{A}) & \!\!\!\!  =  \!\!\!\! &
|\psi_{+}\rangle_{S}\otimes
 \gamma_{0}|0\ldots 0\rangle_{A} + |\psi^{'}\rangle_{S}\otimes
 |\mathbf{j}\rangle_{A} +  \nonumber \\
 {} & \!\!\!\!\!\!\!\!\!\!\!\! \!\!\!\!\!\!\!\!\!\!\!\!\!\!\!\!\!\!\!\! \!\!\!\!\!\!\!\!\!\!\!\!\!\!\!\!\!\!\!\! 
\!\!\!\!\!\!\!\!\!\!\!\!\!\!\!\!\!\!\!\! \!\!\!\!\!\!\!\!\!\!\!\!\!\!\!\!\!\!\!\! \!\!\!\!\!\!\!\!
 + & \!\!\!\!\!\!\!\!\!\!\!\! \!\!\!\!\!\!\!\!\!\!\!\! \!\!\!\!\!\!\!\!\!\!\!\!\!\!\!\!\!\!\!\!\!\!\!\! \!\!  |\psi_{-}\rangle_{S}\otimes
 \sum_{l,m=1}^{s+1}\gamma_{lm}\ad{l}\ad{m}|0\ldots 0\rangle_{A} + |\psi_{--}\rangle_{S}\otimes
 \sum_{l,m,n=1}^{s+1}\gamma_{lmn}\ad{l}\ad{m}\ad{n}|0\ldots 0\rangle_{A}
\end{eqnarray}
\noindent Once again, the transformation can be brought in the
form analyzed in the subsection $2.2$, that is,
$V(U(|\psi\rangle_{S}\otimes |\mathbf{i})) =
U^{''}(|\psi\rangle_{S}\otimes |\mathbf{i}\rangle_{A})$

\section{Conclusions}

In the present paper, we addressed the issue of the maximum
success probability of the post-selected NS-gate. Up to now, this
problem has no general solution: that is, no upper bound which is
both strict and independent of the ancillary resources. Our
strategy was to restrict ancillary resources to only one photon
and arbitrary number of vacuum states. This has reduced the
problem to a mathematically soluble one, still general enough. In
fact, we showed that under these conditions the upper bound is
$0.25$, it is strict, and it is independent of the dimension of
the Hilbert space spanned by the ancillary states. Furthermore, we
have considered generalized post-selection conditions, namely
those described by rank-r projectors, with $r>1$, so that the
problem has been fully solved for the case of a single ancillary
photon.



\end{document}